\documentclass{article}
\usepackage{amsmath}            
\usepackage{epstopdf}           
\usepackage{flushend}           

\usepackage{graphicx}
\usepackage{amssymb}
\usepackage[figuresright]{rotating}
\usepackage{subcaption}
\usepackage[english]{babel}

\usepackage[letterpaper,top=2cm,bottom=2cm,left=3cm,right=3cm,marginparwidth=1.75cm]{geometry}

\usepackage{amsmath}
\usepackage{graphicx}
\usepackage[colorlinks=true, allcolors=blue]{hyperref}

\title{A Novel Framework for Brain Tumor Detection Based on Convolutional Variational Generative Models}
\author{Wessam M. Salama\\Pharos University\\wessam.salama@pua.edu.eg         \and Ahmed Shokry\\ American University in Cairo\\ ahmed.shokry@aucegypt.edu
}
\date{}
\providecommand{\keywords}[1]{\textbf{\textit{Index terms---}} #1}
\begin{document}
\maketitle

\begin{abstract}
Brain tumor detection can make the difference between life and death. Recently, deep learning-based brain tumor detection techniques have gained attention due to their higher performance. However, obtaining the expected performance of such deep learning-based systems requires large amounts of classified images to train the deep models. Obtaining such data is usually boring, time-consuming, and can easily be exposed to human mistakes which hinder the utilization of such deep learning approaches.

This paper introduces a novel framework for brain tumor detection and classification. The basic idea is to generate a large synthetic MRI images dataset that reflects the typical pattern of the brain MRI images from a small class-unbalanced collected dataset. The resulted dataset is then used for training a deep model for detection and classification. Specifically, we employ two types of deep models. The first model is a generative model to capture the distribution of the important features in a set of small class-unbalanced brain MRI images. Then by using this distribution, the generative model can synthesize any number of brain MRI images for each class.  Hence, the system can automatically convert a small unbalanced dataset to a larger balanced one. The second model is the classifier that is trained using the large balanced dataset to detect brain tumors in MRI images. The proposed framework acquires an overall detection accuracy of 96.88\% which highlights the promise of the proposed framework as an accurate low-overhead brain tumor detection system.
\end{abstract}


\section{Introduction}
\label{intro}

Over the years, cancer by its unstable nature remains a curse to humankind \cite{prasad2006magnetic}. 
Computer-aided diagnosis (CAD) applications are used to assist neurologists. Brain tumor detection, classification and grading are presented in  \cite{pan2015brain,deepak2019brain,kumar2017classification,mohan2018mri}. These applications rely on magnetic resonance imaging (MRI) images of the brain, which are better than computed tomography (CT) images because they can provide greater contrast to the soft tissues in the brain compared to CT images. In CAD systems, machine-learning techniques are widely used to detect and classify brain tumors. The basic step for these systems is the feature extraction step where the system learns the important features in MRI images. To this end, several methods of extraction of monuments have been proposed \cite{sajjad2019multi}. The extracted features are then fed into a structured form model to detect and classify brain tumors. However, to make this problem computationally tractable these systems usually assume that the important features in MRI images are independent, which limits their ability to capture the relationship associated with the nature between the features, which in turn reduces their accuracy \cite{shokry2018deeploc}. 

To address these limitations, several brain-based MRI detection systems have adopted deep learning. The convolutional neural network models (CNN) and the transfer learning models are commonly used as solutions for detection and classification problems. These solutions usually contain two phases. The first is an offline phase where the deep model is trained using a set of manually classified MRI images (training data). The second is an online phase that takes a brain MRI image and determines whether it contains a tumor or not.

The limitations of existing solutions are summarized as follows. The performance of modern methods is not sufficient given the medical importance of detection and classification problems. Previous solutions rely on manually defined tumor areas, before classification. This eliminates the disclosure problem and makes the classification problem easier. However, it prevents these systems from being fully automated. On the other hand, automated solutions developed using CNN and its variants have not been able to significantly improve performance. This is because CNN and deep learning models, in general, are data-hungry \cite{rizk2019effectiveness}; i.e. In order to achieve the expected good performance, it requires large amounts of training data (classified images). This comes with additional cost due to the need to perform the assembly and the classification of the brain MRI images. Although pre-trained models (i.e. transfer learning solutions) can be used to solve the need for big training data, the performance of such solutions depend on the pre-training model. If your task (i.e. tumors detection or classification) and the task that the pre-trained model is trained for are too dissimilar, the accuracy will be poor using transfer learning \cite{rosenstein2005transfer}. Finally, existing solutions using either CNN or transfer learning are trained using class-unbalanced datasets (with respect to the number of training samples for each class), lowering their detection and classification accuracy \cite{japkowicz2002class,guo2008class}.

This paper provides a novel general framework for discovering and classifying brain tumors. The proposed framework can be seamlessly integrated with any of the existing MRI-based detection and classification systems by processing their small unbalanced training dataset to produce a larger balanced dataset which is suitable for training deep learning models. In particular, we employ two types of deep models. The first model is an innovative deep model to capture the distribution of the important features in a group of a small class-unbalanced images dataset. Then by using the distribution of the important features, the system can generate any number of brain MRI images for each class. Thus, the system can automatically convert a small unbalanced dataset into a larger balanced one. The second model is the classifier that is trained using the large class-balanced dataset to detect brain tumors in MRI images. 
Note that while we use two models for generating new samples and for detecting tumors, our work is different from the previous work where two models are usually employed for feature extraction task and for classification \cite{gumaei2019hybrid}.

The generator model can provide a dense class-balanced dataset that is required for training the classifier model. This makes it has a large impact on many real-world applications that require dense class-balanced training data. Examples include brain tumor detection, classification, and grading. However, as the classifier model is trained using the newly generated data from the generator model, the accuracy of the classification (i.e. tumor detection) is affected by the accuracy of the generator model. This limitation can be solved if we carefully train the generator model until it reaches certain good accuracy.      

The proposed framework acquires a comprehensive detection accuracy of 96.88\%, outperforming the most recent detection methods, which highlights the promise of the proposed framework as an accurate system for brain tumor detection.

The rest of the paper is organized as follows: Section~\ref{rwork} presents our related work. The proposed framework and the details of the two models are described in Section \ref{method}. The dataset description and the evaluation of the proposed framework are explained in Section \ref{experiments}. Finally, section \ref{conclusion} concludes the paper.

\begin{figure*}[!t]
\centering
\includegraphics[width=1\linewidth,height=3.5cm]{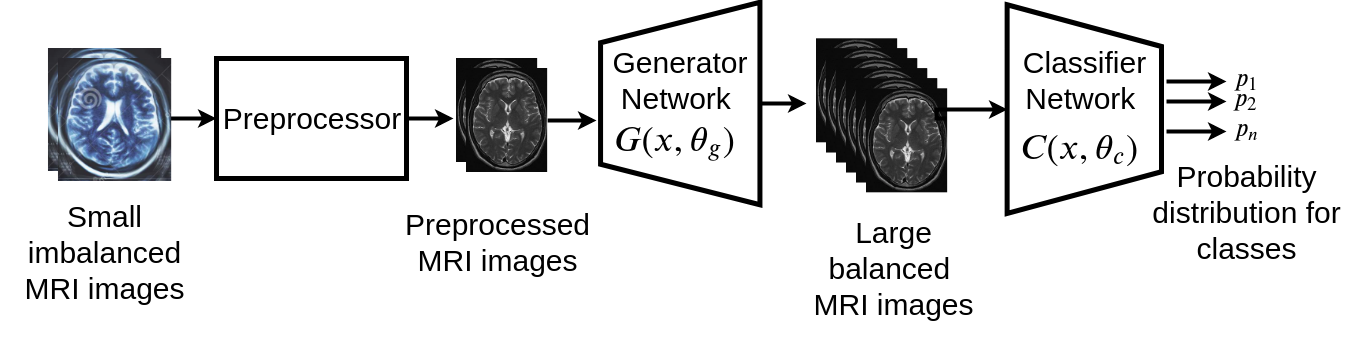}
     \caption{The proposed framework.}
     \label{fig:framework}
\end{figure*}

\section{Related work}
\label{rwork}

\subsection{Deep learning solutions}
Deep learning solutions can learn to generate a high-level feature directly from raw MRI images. Convolutional Neural Networks (CNN) is a commonly used deep model in these systems \cite{lo1995artificial}. It can automatically learn the representation of complex features directly from the data itself. CNN-based brain tumor detection systems have usually two-stage. An offline phase where a deep CNN model is trained using a set of classified MRI images (training data). An online phase that takes a brain MRI image and determines whether it contains tumors or not. CNN-based systems have been successfully applied to the problem of detection and classification of brain tumors. In addition, with the support of parallel GPUs, these technologies have gained tremendous success. 
In \cite{gumaei2019hybrid}, authors further used two different models for developing an accurate brain tumor classification.  The first is a hybrid model for brain tumor feature extraction. The second is a regularized extreme learning machine (RELM) for developing an accurate brain tumor classification. 

\textit{On the other hand, CNN and its variants have not been able to significantly improve performance. This is because CNN and deep learning models, in general, are data-hungry \cite{rizk2019effectiveness}; i.e. In order to achieve the expected good performance, they require large amounts of training data (classified images)}.

\subsection{Transfer learning solutions}
Recently, a special class of deep learning, known as transfer learning, has demonstrated its potential for detecting brain tumors based on MRI \cite{yang2018glioma,jain2019convolutional}. 
Transfer learning allows the use of a previously trained CNN template, which has already been developed for another related application. Several techniques are underway to extract deep features from MRI images of the brain using pre-trained networks. These techniques demonstrate the transfer learning ability to work with smaller datasets \cite{liu2016exploring,ahmed2017fine}.
Examples include \cite{yang2018glioma} where AlexNet and GoogLeNet are employed on grading of glioma from MRI images. In \cite{jain2019convolutional} a pre-trained VGG-16 network is used for the diagnosis of Alzheimer's disease from MRI images.

\textit{On the other hand, the performance of transfer learning solutions depend on the pre-training model. If your task (i.e. tumors detection or classification) and the task that the pre-trained model is trained for are too dissimilar, the accuracy will be poor using transfer learning \cite{rosenstein2005transfer}.}

\subsection{Deep learning-based segmentation}
Recently, there has been much work in image segmentation \cite{zhang2021cross,zhang2020exploring,isensee2021nnu,myronenko20183d,al2021efficient,alzu2020parallel,elbes2019survey,alzu2018transferable,alzu2019multi}. 
In \cite{zhang2021cross,zhang2020exploring} authors propose a novel cross-modality deep-learning based framework to segment brain tumors from the multi-modality MRI data. In \cite{isensee2021nnu}, authors employ nnU-net for brain tumor segmentation. 
In \cite{myronenko20183d}, authors proposed an automated segmentation of brain tumors from 3D MRI images where they used autoencoders to solve the problem of the small training dataset.
Authors of \cite{al2021efficient} also provide an efficient secure algorithm for 3D image segmentation. They proposed an algorithm for segmenting medical volumes based on multiresolution analysis where different 3D volume reconstructed versions have been considered to come up with a robust and accurate segmentation results \cite{alzu2019multi}.

\textit{However, these techniques handle 3D images to perform image segmentation which makes them relatively computationally expensive compared to the use of 2D images which is a better fit for brain tumors detection problem \cite{abiwinanda2019brain,alzu2020parallel,alzu2018transferable}.}

\section{The proposed framework}
\label{method}
Figure~\ref{fig:framework} shows the proposed framework which can be integrated with any of the current deep learning-based brain tumor detection systems\footnote{Without loss of generality, we focus in this paper on the brain tumor detection problem. However, our proposed framework can be applied to other problems such as MRI-based classification and grading.}. The framework takes the brain MRI images data collected by the traditional MRI-based systems as input, typically class-unbalanced and small in size. The input brain MRI images are then passed through the preprocessor module which resizes and normalizes the input brain MRI images. The framework contains two deep models: the Generator model $G(x,\theta_g)$ and the Classifier model $C(x,\theta_c)$ where $x$ is the input to each model, $\theta_g$ is the generator network parameters and $\theta_c$ is the classifier network parameters. The Generator learns the distribution of the important features in the preprocessed images. Then, given the distribution of the important features, the Generator can convert the small unbalanced preprocessed images dataset to a larger balanced one. Finally, the large balanced dataset is used to train another deep model, the classifier, which is used for detection and/or classification. In the balance of this section, we start by describing the input data format followed by the details of the Preprocessor and the Generator modules. The latter is the core contribution of this paper. Finally, we describe the classifier that is used to detect tumors in MRI images.

\subsection{Input data}
The input to the system is a set of brain MRI images. This set is small in size and imbalanced with respect to the number of training images per class. Almost all MRI scanners output images in the standardized medical format. These images are stored as two-dimensional (2D) grayscale images. Each entry in the grayscale image stores a value from 0 to 255. This range presents the trade-off between the efficiency of storing information about the image (256 values fit perfectly in 1 byte) and the sensitivity of the human eye (humans distinguish a limited number of shades of the same color). The grayscale images are then passed through the preprocessor module.

\subsection{Preprocessor}
The goal of this module is to resize and normalize the input MRI images. Firstly, all the input images should be in a fixed size, the grayscale input images are resized to be images of size 256 x 256 pixels. This allows these images to be fed into deep models with fixed input size. Secondly, normalizing the input data generally speeds up learning and leads to faster convergence. To this end, the grayscale input images are normalized to be in the range of $[0,1]$. The normalization is in intensity values. A minimum-maximum normalization technique is followed to scale the intensity values between 0 and 1 \cite{patro2015normalization}. The preprocessed images are then passed through the generator network.

\begin{figure*}[!t]
\centering
\includegraphics[width=1\linewidth,height=4.5cm]{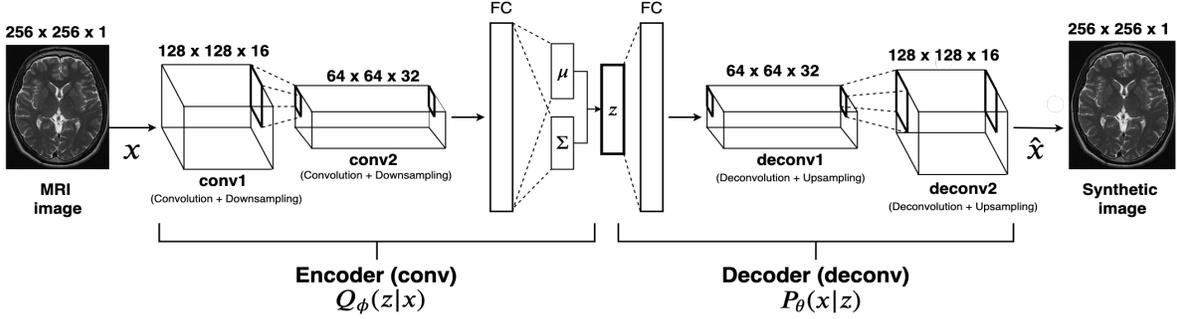}
     \caption{The generative network structure. Input image of size 256 x 256 is passed through the encoder network $Q_{\phi}(z|x)$ which has two convolutional and maxpooling layers to extract the most important features from the input image $x$. The output from the encoder is transformed to one vector (i.e. flattened) then passed through a fully connected layer (FC) to get the mean $\mu$ and the standard deviation $\Sigma$ of the encoder distribution $Q_{\phi}(z|x)$. We then sample $z$ from the $Q_{\phi}(z|x)$ and pass it through the decoder network $P_{\theta}(x|z)$ with two deconvolutional and upsampling (i.e. nearest neighbors) layers to reconstruct the image $\hat{x}$. We force the generative network to reconstruct the input images while making $z$ follows the normal distribution.}
     \label{fig:cvae}
\end{figure*}

\subsection{Generator model}
\label{generatorr}
The goal of the generative model is to synthesize new brain MRI image samples for each class. The generative model is trained using the preprocessed images to generate new image samples that reflect the typical patterns in the preprocessed images.   
Our generative model is a modified version of the variational autoencoder \cite{kingma2013auto}.
In general, autoencoders are a specific type of feedforward neural networks where the input is the same as the output \cite{baldi2012autoencoders}. They are used to learn key representational information (features) for a dataset within a low-dimensional latent space in an unsupervised mannar where they compress the input image into a lower-dimensional latent-space representation $z$ (embeddings) and then reconstruct the output image from this representation. Hence, the latent space learns to capture the most essential information required for reconstruction. However, the latent space embeddings may be sparsely distributed that makes the key information to be spread across several clusters in the latent space. Meanwhile, the empty space between clusters does not capture any useful information which makes sampling from it creates meaningless results.

To solve this problem, we uses the variational autoencoder (VAE) where a new constraint is added that the latent space embeddings need to follow certain predefined distribution $p(z)$ \cite{kingma2013auto}. This distribution is usually selected as normal distribution \cite{doersch2016tutorial}. Now, by forcing the latent space embeddings to follow the normal distribution, the network is forced to fully utilize the latent space so that information is distributed in a way that allows us to sample from any point in the latent space to generate new images that reflect the typical patterns in the original small brain MRI images dataset. Therefore, we depend on VAE to generate new brain images.

Because brain tumors do not always appear with the same number, same shape and in the exact same position in the brain MRI images, convolutional and deconvolutional layers are applied in the implementation of the encoder and the decoder instead of the regular feedforward layers. These layers can utilize sliding filter maps that can recognize the tumors' local patterns independently of their number, shapes, and positions in the brain MRI images. Hence, the generator model learns to generate new MRI images with different numbers of tumors, with different shapes, and in different positions in the images. Figure~\ref{fig:cvae} shows our convolutional variational autoencoder (CVAE) architecture. The CVAE network has two main components: the encoder (conv) and the decoder (deconv). The encoder network consists of several convolutional layers followed by a fully connected layer and the decoder network consists of a fully connected layer followed by convolutional layers. 
The encoder compresses the brain MRI image input $x \in X$ to get the hidden (latent) representation $z$ and network parameters $\phi$ as an output. The latent space $z$ is typically referred to as a \textit{bottleneck} because the encoder must learn an efficient compression of the brain image data into a lower-dimensional space. 
We refer to the encoder as $Q_{\phi}(z|x)$.
We can sample from this distribution to get noisy values of the representations $z$. The decoder takes the latent representation $z$ as input and produces the parameters to the probability distribution of the data and has weights and biases $\theta$. The decoder is denoted by $P_{\theta}(x|z)$. The loss function is the negative log-likelihood function defined as follows,

\newcommand{\Lagr}{\mathcal{L}}
\begin{equation}
 L(\phi, \theta) = \Lagr_1 + \Lagr_2
\end{equation}
\begin{equation}
\Lagr_1 = - E_{z\sim Q_\phi(z\vert x)}[\log P_\theta(x\vert z)]
\end{equation}
\begin{equation}
\Lagr_2 = KL(Q_\phi(z\vert x) \vert\vert p(z))
\end{equation}

The loss consists of two terms. The first term is the reconstruction loss. It is the expected negative log-likelihood of the data. This term forces the decoder to learn to reconstruct the data which prevents the VAE from generating meaningless image samples. The second term is the regularizer. It is the Kullback-Leibler \cite{joyce2011kullback} divergence between the encoder's distribution $Q_\phi(z\vert x_i)$ and the predefined distribution $p(z)$. Assuming that $p(z) = N(0,1)$, the regularizer forces the latent representations $z$ to follow the standard normal distribution. We employ the stochastic gradient descent to optimize the loss with respect to CVAE network parameters $\phi$ and $\theta$.

We employ convolutional variational autoencoder (CVAE) model for each class in the dataset to learn the joint distribution $P(X_i)$ of input features over the small training images $X_i$ for class $i$\footnote{Note that in our detection problem we have two classes: images have tumors or not. Hence, we used two identical CVAE models. One is trained with "Yes" images and another one is trained with "No" samples.}. After finishing the training phase, the network can generate new brain MRI images by sampling the latent variables $z \sim N(0,1)$, then decode $z$ to get new brain image samples $\hat{x}$ from class $i$. So, we can convert a small unbalanced dataset to a larger balanced one.

\subsection{Classifier model}
\label{classifierr}
This section describes our classifier model and how we increase the model robustness.

\begin{figure*}[!t]
\centering
\includegraphics[width=1\linewidth,height=3.2cm]{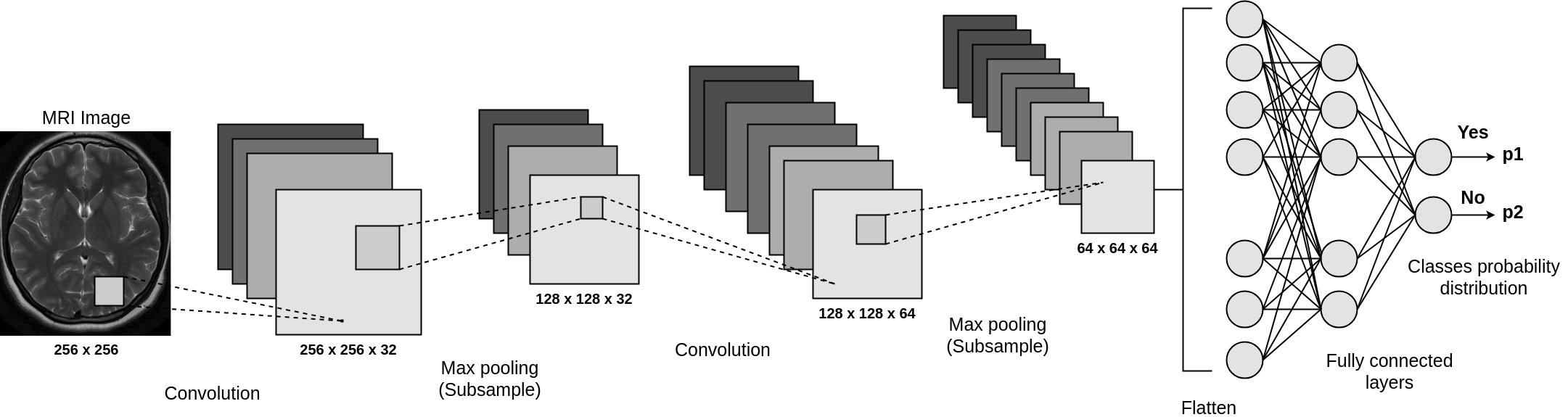}
     \caption{The classifier structure. Input image of size 256 x 256 is fed into the classifier. The convolutional layer uses 32 filter to generate the features map. The image is then passed through the maxpooling layer which downsample image dimentions by a factor of 2. This process is repeated by the second convolutional and pooling layers. After that, we flatten the output of the second pooling layer to transform the output to one vector. Finally, a fully connected layer followed by a softmax layer are used to get the classes probability distribution.}
     \label{fig:cnn}
\end{figure*}

\subsubsection{Basic model}
Our classifier structure is represented in Figure~\ref{fig:cnn}. The classifier model is assumed to be a binary classifier that detects whether a brain MRI image has a tumor or not. However, this model can be replaced with any other classifier. The classifier model is a convolutional neural network (CNN). The input to the CNN is a brain MRI image. The output is the probability distribution for the different classes. The classifier consists of three components. The first one is the convolutional layers which extract the most important features from the brain MRI image. The second is the pooling layers that downsample each feature to reduce its dimensionality and focus on the most important elements. Finally, the fully connected layers are used after flattening the features (that identified in the previous layers into a vector) to predict the probability that the brain MRI image has tumors or not.
\subsubsection{Increasing Model Robustness}
To further increase the model robustness, the proposed system employs the drop-out regularization technique during training \cite{srivastava2014dropout}. The
idea is to randomly drop neurons from the network during training. The temporarily removed neurons no
longer contribute to the activation of downstream neurons in the forward pass. Similarly, the weight update process is not applied to them in the backward pass. This prevents the network from overfitting the training data.

\begin{figure*}[!t]
        \centering
        \begin{subfigure}[b]{0.24\linewidth}
                \includegraphics[width=0.95\linewidth]{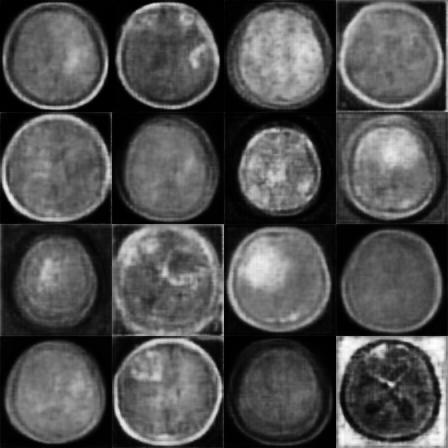}
           \caption{Iteration 100.}
           \label{}
        \end{subfigure}%
        \begin{subfigure}[b]{0.24\linewidth}
                \includegraphics[width=0.95\linewidth]{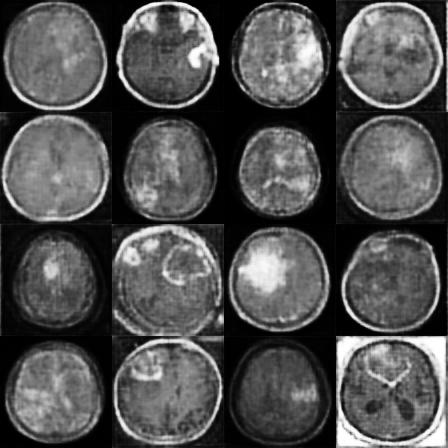}
              \caption{Iteration 300.}
            \label{}
        \end{subfigure}
        \begin{subfigure}[b]{0.24\linewidth}
                \includegraphics[width=0.95\linewidth]{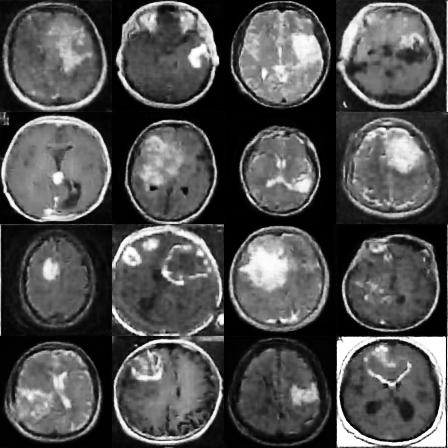}
           \caption{Iteration 1700.}
           \label{}
        \end{subfigure}%
        \begin{subfigure}[b]{0.24\linewidth}
                \includegraphics[width=0.95\linewidth]{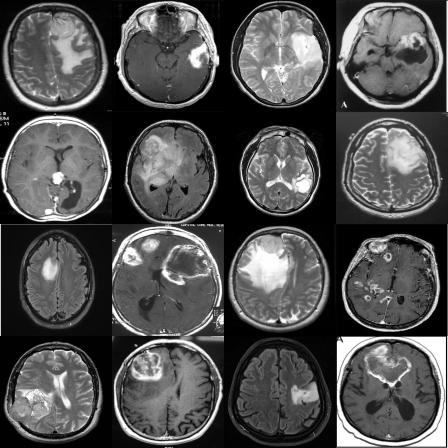}
              \caption{Iteration 2000.}
            \label{}
        \end{subfigure}        
        \caption{Synthetic brain MRI images. Initially at iteration 100 from the training of generator network, the generator produces vague brain images. As more we increase training iterations, as more we get clearer brain images that reflect the typical patterns in the brain MRI images.}
             \label{fig:gen_per}
\end{figure*}
\begin{figure*}[!t]
        \centering
        \begin{subfigure}[b]{0.33\linewidth}
                \includegraphics[width=\linewidth]{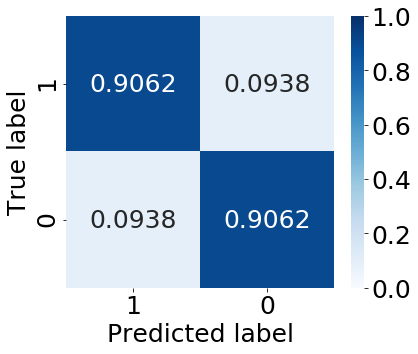}
           \caption{CNN classifier.}
           \label{}
        \end{subfigure}%
        \begin{subfigure}[b]{0.33\linewidth}
                \includegraphics[width=\linewidth]{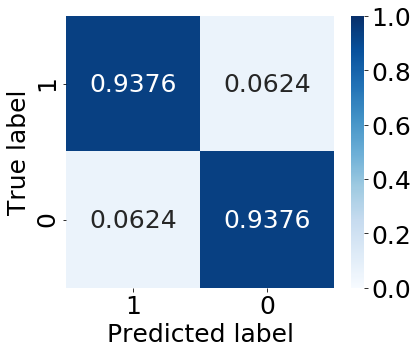}
           \caption{Transfer learning classifier.}
           \label{}
        \end{subfigure}%
        \begin{subfigure}[b]{0.33\linewidth}
                \includegraphics[width=\linewidth]{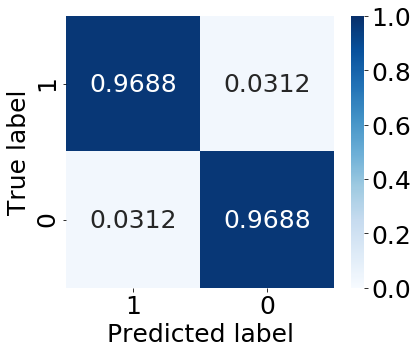}
           \caption{Proposed framework.}
           \label{}
        \end{subfigure}%
	\caption{Normalized confusion matrix for different approaches.}
	\label{fig:cpf}
\end{figure*}

\subsection{Discussion}
The proposed framework consists of two different models for two different functions. The generator model target is to generate more data samples for each class in a small class-unbalanced data. On the other hand, the classifier model target is used only for classification. The generator model needs to be trained first to generate the synthetic samples. The original samples and the synthetic ones are then used for training the classifier. Hence, the two networks cannot be trained in an end-to-end manner. 

The generator network can be employed independently from the classifier network. It can be used to generate more data samples for any image-based machine learning problem that needs a dense balanced dataset for training. The classifier, on the other hand, is used for any image-based classification. Hence, our framework is general for different detection and classification problems as we can replace the classifier of the brain tumor detection with another one that handles another problem (e.g. grading the tumors).

\section{Experiments and results}
\label{experiments}
This section starts with dataset description, followed by the explanation of the performance metrics. Finally, we compare the performance of the proposed framework with other systems. We implement the generator and the classification networks using Google Colab.

\subsection{Dataset description}
We depend in our experiments on a small dataset that contains a few number of samples, typically 253 samples. The dataset is openly available in \cite{dskey} and is commonly used for evaluating classification in \cite{thotapally2020brain}. Each sample in the dataset has a binary label that indicates whether the brain MRI image sample has a tumor or not. We used 70\% of the samples for training and the remaining 30\% for testing.\\ 

\subsection{Framework settings}
The performance of the brain tumor classification framework depends on a combination of generator model parameters and classifier model parameters. There are three distinct framework settings.

\subsubsection{Deep CNN model as a stand-alone system (CNN)}
The deep convolutional neural network is used as a classification model. The network is trained using the small training dataset (after pre-processing). The network architecture is described in section \ref{classifierr}. The hyperparameters of the network are heuristically adjusted. RMSprop optimizer is chosen as an optimizer. A high value for the learning rate makes the optimizer overshoots the minimum value for the loss, while a small value for the learning rate increases the training time. We choose the learning rate to be 0.0001. The loss function is categorical cross-entropy which measures the closeness of the predicted and actual distributions. The dropout rate is 0.5. Table \ref{gen_clas_param} summarizes the classifier parameters.

\subsubsection{Transfer learning}
A pre-trained modified VGG network followed by a dense layer with 256 neurons and the output layer with two neurons is trained using the small training dataset (after pre-processing). RMSprop is used as an optimizer with a learning rate of 0.0004. The loss function is categorical cross-entropy. The dropout rate is chosen to be 0.5.

\subsubsection{Proposed framework}
Our proposed framework with two networks: the generator network and the classifier. The preprocessed images are used to train the generator network described in section \ref{generatorr}. Figure~\ref{fig:gen_per} shows a sample of generated images using the generative network. In the start of training the generative network, the network was not able to generate clear brain MRI images. With increasing the training time (i.e.number of training iterations), the network generates clearer images that eventually reflect the typical pattern of the brain MRI images. We used the generator network to convert the original small training dataset to a larger one with 1000 samples (500 for each class). The newly generated large dataset is then used for training the classifier network for detection. The classifier network is described in section \ref{classifierr}. We tried different architectures for the generator and the classifier networks. Table \ref{gen_clas_param} contains the generator and the classifier best parameters in terms of the detection accuracy.\\

\begin{table}[!t]
	\centering
	\caption{Generator and classifier networks parameters}
	\label{gen_clas_param}
	\scalebox{0.99}{
		\begin{tabular}{|l||l|p{2cm}|}\hline
			\textbf{Parameter} & \textbf{Generator} & \textbf{Classifier}\\ \hline \hline
            Architecture & Fig\ref{fig:cvae} & Fig\ref{fig:cnn} \\ \hline			
			Learning rate & 0.0001 & 0.0001 \\ \hline
			Batch size & 16 & 32\\ \hline
			Dropout rate & 0.1 & 0.5 \\ \hline
			Number of training epochs & 2000 & 100\\ \hline
			Size of input layer & 256x256 & 256x256 \\ \hline
			Size of output layer & 256x256 & 2 \\ \hline
		\end{tabular}
	}
\end{table}

\subsection{Performance metrics}
There are several evaluation tools to assess a classifier amongst them, are the accuracy, the precision, the recall, and the F1 score. Accuracy is the measure of a correct prediction made by the classifier. It gives the ability of the performance of the whole classifier. The accuracy is defined as,

\begin{equation}
Accuracy = \frac{TP + TN}{TN + FP + FN + TP}
\end{equation}
Precision, on the other hand, is the ratio of correctly predicted positive observations to the total predicted positive observations. High precision relates to the low False Positive Recall (FPR):
\begin{equation}
Precision = \frac{TP}{TP + FP} 
\end{equation}
Recall in this context is also referred to as the true positive rate
\begin{equation}
Recall = \frac {TP}{TP + FN}
\end{equation}
The F1-score is the weighted average of precision and recall. It is used as a statistical measure to rate the performance of the classifier. Therefore, this score takes both false positives and false negatives into account:  
\begin{equation}
   F_1=\frac{2*(Recall * Precision)}{(Recall + Precision)} 
\end{equation}

\subsection{Model accuracy}
\begin{figure}[!t]
        \centering
        \begin{subfigure}[b]{0.5\linewidth}
                \includegraphics[width=0.98\linewidth]{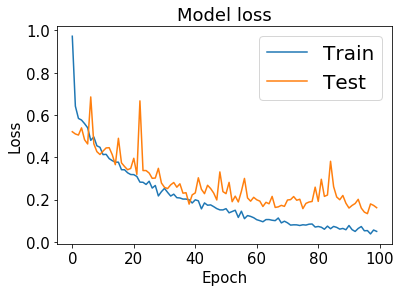}
           \caption{Loss of the classifier.}
           \label{}
        \end{subfigure}%
        \begin{subfigure}[b]{0.5\linewidth}
                \includegraphics[width=0.98\linewidth]{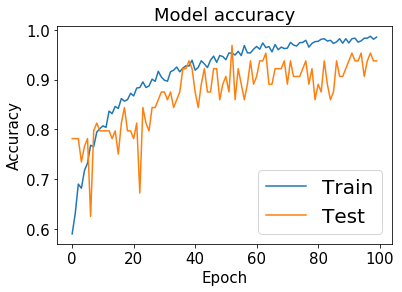}
              \caption{Accuracy of the classifier.}
            \label{}
        \end{subfigure}      
        \caption{Loss and accuracy.}
             \label{fig:loss_acc}
\end{figure}

Figure \ref{fig:loss_acc} shows the loss and detection accuracy for training and testing for the classifier network with an increasing number of training epochs. Evidence from figure \ref{fig:loss_acc} that the detection accuracy for training and testing increases together. It also shows that the loss of training and testing decreases together. This indicates that the classifier network does not over-train (over-fit) the training data. This is mainly because the newly generated variational samples help in generalizing the classifier.\\
\subsection{Effect of number of generated samples}
Figure~\ref{fig:nsamples} shows the effect of increasing the number of additional generated training samples on the system accuracy. The figure shows that, as expected, increasing number of training samples increases the system accuracy due to the increase of the quality of the learned model. The good news is that the system performance saturates after about 2000 samples.

\subsection{Comparison with other systems}
Now, we compare our proposed framework with the CNN classifier model and the transfer learning model both in accuracy and training time.
\subsubsection{Accuracy comparison}
We compare the proposed framework with other approaches such as training the classifier network directly using the original training data (CNN) and training the pre-trained VGG network directly using the original training data (Transfer learning). Figure \ref{fig:cpf} shows the normalized confusion matrix for the different approaches.\\

\begin{table}[!t]
	\centering
	\caption{Summary of performance metrics for the different classifiers.}
	\label{perf_m}
	\scalebox{1}{
		\begin{tabular}{|l||l|l|l|p{2cm}|}\hline
			System & Accuracy & Precision & Recall & F1-score\\ \hline \hline
			CNN  & 90.62\% & 90.62\%  & 90.62\% & 90.62\% \\ \hline
			Transfer learning & 93.75\% & 93.75\%  & 93.75\% & 93.75\% \\ \hline
			\textbf{Proposed framework} & \textbf{96.88\%} & \textbf{96.88\%}  & \textbf{96.88\%} & \textbf{96.88\%} \\ \hline
		\end{tabular}
	}
\end{table}
Table \ref{perf_m} summarizes the performance metrics such as Accuracy, Precision, Recall, and F1-score for the different approaches. The results show that the different approaches can classify the positive samples (i.e. the samples that have a tumor) as good as the negative samples. The transfer learning technique is better than the CNN classifier as it depends on a large number of pre-trained weights. Finally, the table shows that our proposed framework can detect tumors better than other approaches. This is because of the ability of the generative model to augment the CNN with newly generated samples which highlights the promise of the proposed framework as an accurate low-overhead brain tumor detection system.

\begin{figure*}[!t]
        \centering
                \begin{minipage}{0.48\linewidth}
                \includegraphics[width=0.98\linewidth]{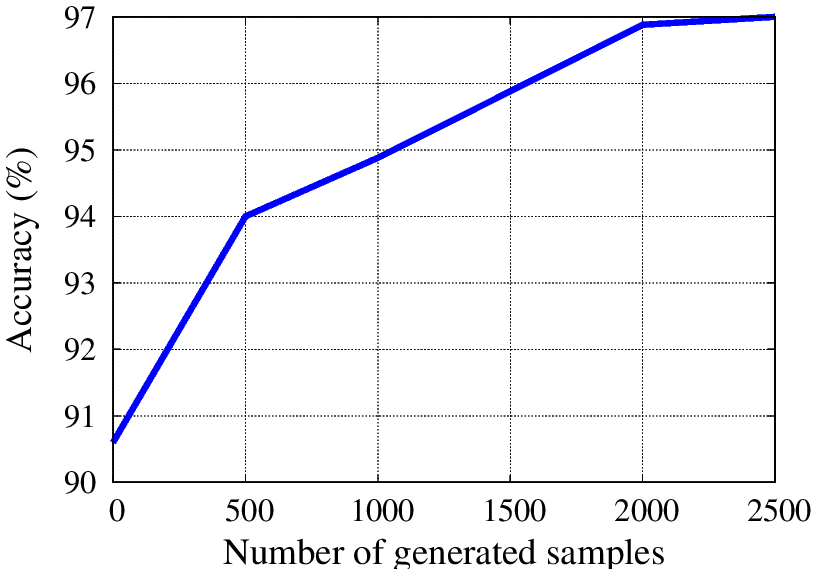}
              \caption{Effect of number of generated\\ samples on the accuracy.}
            \label{fig:nsamples}
        \end{minipage}
        \begin{minipage}{0.48\linewidth}
                \includegraphics[width=0.98\linewidth]{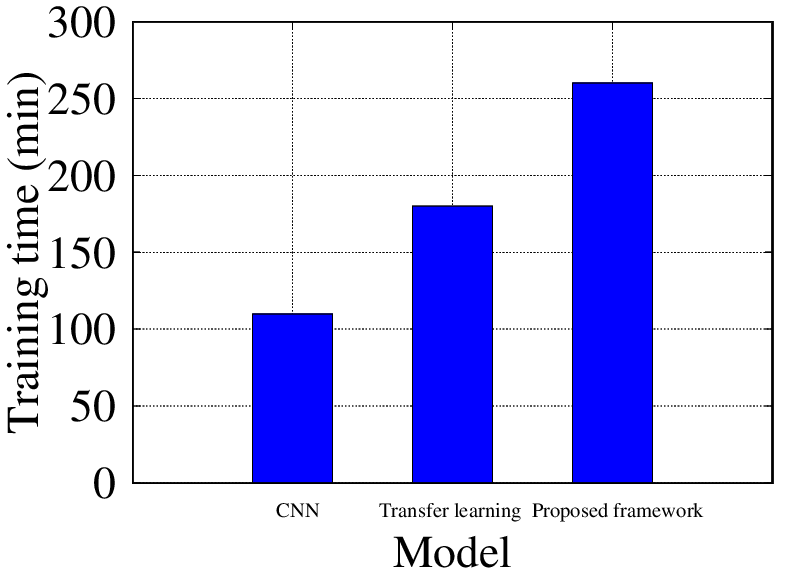}
           \caption{Training time for the different techniques.}
           \label{fig:time}
        \end{minipage}%
\end{figure*}

\subsubsection{Time comparison}
Figure~\ref{fig:time} shows the training time for the three models. The CNN classifier model training is the fastest as it contains few layers. Unlike the CNN classifier, the transfer learning classifier contains a relatively large number of layers to train. Finally, as our proposed network trains the generative model before training the CNN classifier, our proposed framework is the slowest. Note that, the training is achieved in an offline stage. Hence, it doesn't affect the models' online running time where the three models are fast.

\section{Conclusion}
\label{conclusion}
A novel generalized framework for brain tumors detection and classification is introduced in this paper. The proposed framework used two different deep models for two different tasks. The first is a convolutional variational generative model to convert a small class-unbalanced dataset to a large balanced one. The second is the classifier which is a convolutional model used to detect tumors in brain MRI images. The proposed framework acquired an equal best performance of accuracy, precision, recall, and F1-score of 96.88\%, outperforming other recent systems in the literature. This highlights the promise of our framework as an accurate brain tumor detection system.\\

\bibliographystyle{plain}
\bibliography{sample}

\end{document}